# Report of the International Committee for Future Accelerators‡

**Pushpalatha C. Bhat**[1]
*Fermi National Accelerator Laboratory*
*P. O. Box 500, Batavia IL 60510, USA*
*E-mail:* `pushpa@fnal.gov`

**Geoffrey N. Taylor**[2]
*Centre of Excellence for Particle Physics at the Terascale*
*University of Melbourne, Victoria 3010, Australia*
*E-mail:* `gntaylor@unimelb.edu.au`

The International Committee for Future Accelerators (ICFA) has been in existence for well over four decades. Its mission is to facilitate international collaboration in the construction and use of accelerators for high energy physics. This report presents, after a brief introduction, some recent activities of ICFA and its panels. The International Linear Collider (ILC) and its current status are briefly discussed.

---

[‡] http://icfa.fnal.gov/;  [1] ICFA Secretary ; [2] ICFA Chair

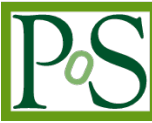

1.  Introduction

IUPAP's Commission on Particle Physics (C11) created the International Committee for Future Accelerators (ICFA) [1] as its first working group in 1976 pursuant to the recommendation of leaders of the particle physics community from around the world who met for a key international meeting in New Orleans in 1975. ICFA's mandate consists of (1) promoting international collaboration in the construction and exploitation of large accelerators for high energy physics (HEP); (2) organizing regularly world-inclusive meetings for the exchange of information on future plans for regional facilities and for the formulation of advice on joint studies and uses; and (3) organizing workshops for the study of problems related to super high energy accelerator complexes and their international exploitation and to foster research and development of necessary technology.

Given its origins, ICFA is the recognized body representing the world particle physics community on the global stage. Its membership is representative of particle physics activity in different regions of the world. It includes directors of major particle physics laboratories and other regional representatives. The chair of ICFA is chosen by an internal committee (for a term of three years) and the position rotates between Americas, Europe and Asia. The chair of IUPAP C11 serves as an ex-officio member of ICFA. More details can be found at the ICFA website http://icfa.fnal.gov/.

2.  Recent Activities of ICFA and its Panels

Generally two meetings are held each year, one of which is the annual two-day meeting in the winter (February/March) and the other is a one-day meeting held during ICHEP or the Lepton-Photon conferences which alternate annually. Every third year, a Seminar with the title, "Future Perspectives in High Energy Physics", that runs for multiple days (generally three and a half days), is held. These triennial Seminars feature invited talks on the status and prospects of accelerators, particle physics and related fields.

The 2018 annual winter meeting of ICFA was held March 8-9, 2018 at the Emmanuel College, University of Cambridge, U.K. Prof. Mark Thomson of U. Cambridge was the official host and the meeting was co-sponsored by Fermilab and Emmanuel College. As is the tradition for these longer annual meetings, in addition to the ICFA members and panel Chairs and special guests, the directors of several laboratories conducting particle physics and/or accelerator physics research were also invited to have a broad discussion on the activities and advances in the field. As part of this meeting, the Linear Collider Board (LCB), one of the Panels of ICFA (more details below) also met, as did the Funding Agencies for Large Colliders (FALC), chaired by Grahame Blair of STFC and Oxford. The one-day (summer) meeting was held on July 8, 2018 at this ICHEP meeting in Seoul, Korea. This meeting also had brief sessions of the LCB and FALC.

The most recent triennial Seminar took place November 6-9, 2017 in Ottawa, Canada. The program committee for the Seminar in Ottawa was chaired by Jonathan Bagger, the Director of TRIUMF, and comprised over a dozen other expert physicists from across the particle, accelerator and astroparticle physics communities, some of whom were ICFA members. The Seminar covered topics of neutrinos, dark matter, QCD, Higgs and Electroweak, flavor physics as well as accelerators, detectors, medical isotopes, data science and new technologies. The committee strived for regional and gender balance and achieved it with excellent results and very high quality. There were 212 registered participants – 92 from Americas, 78 from Europe and 42 from Asia; 44% of speakers were female.

The Seminar was also very successful in attracting government officials to the meeting. Honorable Julie Payette, the Governor General of Canada attended the reception hosted at the Canadian Museum of History. Governor General Payette addressed the attendees along with individual conversations. In the closing session of the meeting, a keynote address was given by the Parliamentary Secretary of Science in the Government of Canada, Kate Young, and a round table discussion with some of the leaders in our field, moderated by Paul Wells of Maclean's magazine was held. The meeting also had an embedded popular science communicator, Ian O'Neil of Astroengine, who spoke about his impressions of the meeting. The Seminar was highly informative, productive and enjoyed by the participants.

Over time, ICFA has set up several Panels, each consisting of approximately sixteen experts on the specific technical areas relevant to the Panel. Each organizes its own program, including workshops and schools, newsletters, and collaborative R&D. Each Panel also regularly reports at the ICFA meetings. ICFA recently approved new "Policies and Procedures" providing guidelines for the management of the Panels. The currently active Panels are (Chair of the Panel indicated in parentheses):

- Linear Collider Board (T. Nakada, Lausanne)
- Beam Dynamics (Y. Chin, KEK)
- Instrumentation Innovation and Development (A. Cattai, CERN)
- Advanced and Novel Accelerators (B. Cros, CNRS)
- Interregional Connectivity (H. Newman, Caltech)
- Data Preservation in HEP (C. Diaconu, CPPM, Marseille)
- Sustainable Accelerators and Colliders (M. Seidel, PSI)

A Panel on neutrinos, chaired by K. Long, recently completed its study and produced a roadmap document [2].

Examples of some recent activities of Panels are outlined here. The Instrumentation Innovation & Development (II&D) Panel periodically holds two kinds of instrumentation schools around the world: EDIT (Excellence in Detector Instrumentation and Technology) schools that are aimed at young scientists (graduate students and postdocs) working in collaborations at major particle physics laboratories; and another series in developing countries to motivate and train students from disadvantaged parts of the world. The latter are supported from funds and in-kind contributions of instrumentation from major labs, used at the schools and instrumentation gifted to the hosting institution. The latest such school was held in Havana, Cuba from November 27 – December 8, 2017. An EDIT school was held at Fermilab during March 5-16, 2018. The Beam Dynamics Panel sponsored and organized a High Brightness Hadron Beams workshop (HB2018), one of its highly successful workshop series, in Daejeon, South Korea, June 17-22, 2018, attended by approximately 150 participants. The Beam Dynamics Panel also publishes a popular newsletter twice or thrice a year, which contain comprehensive review articles on recent developments in relevant technical areas. The Panel on Advanced and Novel Accelerators (ANA) held several meetings and workshops in the past year to discuss the roadmap for design of future facilities that would use advanced acceleration techniques. The Sustainable Accelerators and Colliders Panel held its fourth workshop in Maguerele, Romania in November 2017.

3.     ICFA and The International Linear Collider (ILC)

Since ~2000, ICFA has been actively engaged in efforts, on behalf of the world particle physics community, to enable the realization of a linear electron-positron collider. ICFA created the International Linear Collider Steering Committee (ILCSC) in 2002 to promote the construction of an electron-positron collider through worldwide collaboration and created the International Technology recommendation Panel (ITRP) in 2003 that then recommended the choice of superconducting radiofrequency (SRF) technology for the ILC. Subsequently, in 2005, ICFA set up the Global Design Effort (GDE) with Barry Barish as the Director with the charge to produce an ILC design and cost estimates. The GDE was hosted at Fermilab. In 2013, the GDE released the completed Technical Design Report (TDR) for an ILC with a center of mass energy of 500 GeV. The ILCSC was terminated at that point and the Linear Collider Board (LCB) was formed as a Panel of ICFA. The LCB was given the mandate to oversee the work on the ILC as well as to bring under the same umbrella the Compact Linear Collider (CLIC) project efforts. The Linear Collider Collaboration (LCC) was formed to facilitate the R&D and exploit the synergies between the ILC and CLIC efforts. Lyn Evans was appointed as the LCC Director. The initial three-year mandate of the LCB was updated by ICFA in 2016, extending it through 2019.

The monumental discovery of the Higgs boson [4,5] with a mass around 125 GeV at the LHC in 2012, re-energized the global HEP community and began to influence the ILC aspirations and project goals. The Japanese HEP community expressed interest in building the ILC in Japan. Since the Higgs boson mass was found to be 125 GeV, staging of the ILC was foreseen, starting with a 250 GeV machine (a Higgs factory) to study the Higgs physics in the clean $e^+e^-$ environment with great precision. With the possibility of subsequent upgrades to 350 GeV (top-antitop threshold) and 500 GeV, a staged ILC became a very attractive option. The Linear Collider Collaboration (see http://www.linearcollider.org) carried out extensive studies and R&D on accelerator, detector and physics issues for the past several years. Various staging options and costing scenarios have been studied. The LCB has been working closely with ICFA and has been holding extensive meetings in conjunction with the ICFA meetings for the past few years where the results of the studies and R&D are reported and discussed. At the ICFA Seminar in November 2017, the LCB reported to ICFA its conclusions on the 250 GeV ILC as Higgs Factory proposed by the Japanese HEP (JAHEP) community. The conclusions strongly supported the JAHEP proposal to construct the 250 GeV ILC in Japan and encouraged the Japanese government to give the proposal serious consideration in a timely decision. At the same meeting in November 2017, ICFA endorsed the LCB conclusions and issued a Statement [6] very strongly encouraging Japan "to realize the ILC in a timely fashion as a Higgs boson factory with a center of mass energy of 250 GeV as an international project, led by the Japanese initiative."

It is also important to note that there has been strong international support for the ILC since the discovery of the Higgs boson. The 2013 European Strategy for Particle Physics report [7] states, "the initiative from the Japanese particle physics community to host the ILC in Japan is most welcome, and European groups are eager to participate. Europe looks forward to a proposal from Japan to discuss a possible participation." The 2014 United States Particle Physics Project Prioritization Panel (P5) [8] urged the need to use the Higgs boson as a new tool for discovery, and noted, "Motivated by the strong scientific importance of the ILC and the recent initiative in Japan to host it, the U.S. should engage in modest and appropriate levels of ILC accel-erator and detector design in areas where the U.S. can contribute critical expertise. Consider higher levels of collaboration if ILC proceeds."

The LCC studies of the staging and cost reduction exercises [9] indicate that a 250 GeV ILC will be approximately 40% lower in cost than a 500 GeV ILC of the original TDR. The cost reduction comes mainly from the smaller length of the accelerator and latest advances in SRF technology [10].

4.     The Case for the 250 GeV ILC

The discovery of the Higgs boson in 2012 significantly strengthened the physics case for the ILC. With the mass of the Higgs boson being 125 GeV, a 250 GeV ILC would serve as a Higgs factory that would enable precision Higgs measurements, measuring couplings to Standard Model particles at about a percent level or better. From the technology point of view, the ILC design is very mature; the SRF R&D program at Fermilab and elsewhere has delivered accelerating gradients in excess of the ILC specification of 31.5 MV/m and an order of magnitude better RF cavity quality factor than what was foreseen. The cost reduction of ~40% for a 250 GeV ILC relative to the 500 GeV ILC of the TDR design is a major factor in the keenness and optimism in the recent efforts towards its realization. Being a linear machine, the ILC would be upgradeable to higher energies, to the top-antitop threshold and beyond, with the addition of accelerating modules.

The European XFEL at the DESY Laboratory in Germany, which has been now commissioned and operating with 17.5 GeV $e^-$ beam can be, in fact, considered as a large-scale prototype of the ILC. It has 101 cryogenic modules with SRF cavities providing 23.6 MV/m of accelerating gradient and with accelerator length of 2.1 km.

Extensive physics studies by the LCC [11] and other working groups (such JAHEP) [12] have shown that the 250 GeV ILC will provide excellent science from precision studies of the Higgs boson and that the ILC has the potential to discover new particles from beyond the standard model (BSM) physics. Therefore, ICFA considers the ILC as a "key science project complementary to the LHC and its upgrade".

5.     The ILC Status

The ILC proposal by the JAHEP is under serious consideration by the Japanese government. Since the European particle physics community has now started its new round of Particle Physics Strategy Update exercise [13] and has requested input on all international future projects and proposals by December 18, 2018. ICFA has, therefore, proposed that the decision on hosting the ILC by Japan be made by the end of 2018. The Japan Ministry of Education, Culture, Sports, Science and Technology (MEXT) has had several working groups meeting throughout the year. A high level ILC advisory panel of MEXT has also had a series of meetings and deliberations on all aspects of the ILC. The advisory panel made its recommendation [14] and has asked the Science Council of Japan (SCJ) to review the project. At the time of this meeting, the SCJ was preparing to conduct extensive reviews and prepare its final report on the timescale of a few months to enable a decision by the Japanese government by the end of the year.

There has also been significant outreach both domestic and abroad by Japanese organizations and agencies to promote the ILC in Japan. There have been meetings on the ILC between Prime Minister Abe, MEXT officials and physics community representatives. Japanese officials have also had encouraging visits and interactions with government agencies in the US and Europe. There have been strong on-going efforts in Japan with outreach to the public, media, broader science community and Industry.

6.      Summary and Conclusions

Particle Physics research is a global enterprise. International collaborations in HEP are thriving in Europe, Americas and Asia, making discoveries and producing a rich harvest of physics, as seen from the presentations at this conference. The ICFA global strategy is to promote and facilitate international collaboration and coordination in planning future large accelerator facilities, providing regional balance and global benefits. It is guided by three basic requirements: (1) physics drivers, (2) technology, and (3) resources, with the goal of advancing particle physics at the energy and intensity frontiers. At the energy frontier, the key focus beyond the LHC and HL-LHC projects is currently the ILC in Japan. In addition, ICFA anticipates to have extensive deliberations on other energy frontier project proposals in the near future, such as CEPC/SPPC in China, HE-LHC, CLIC and FCC-ee and FCC-pp at CERN. The Long Baseline Neutrino Facility (LBNF) project in the U.S. and J-Parc in Japan are the major neutrino programs at the intensity frontier.

For now, the global particle physics community awaits the most consequential decision on the ILC in Japan while ICFA continues to champion the cause.

**Acknowledgements**

Authors thank the ICFA and LCB members and ICFA Panel chairs for their collaboration and input. Authors also thank the ICHEP2018 organizing committee for a wonderful conference. PB is supported by Fermi National Accelerator Laboratory (Fermilab), which is managed by the Fermi Research Alliance, LLC (FRA) under the contract #DE-AC02-07CH11359 with the U.S. Department of Energy. GT is supported by the Australian Research Council and the University of Melbourne.